\documentclass[aps,prb,twocolumn,showpacs,floatfix]{revtex4}
\usepackage{graphicx}
\usepackage{dcolumn}
\usepackage{amsmath}

\newcommand{\be}{\begin{equation}}
\newcommand{\ee}{\end{equation}}
\newcommand{\bea}{\begin{eqnarray}}
\newcommand{\eea}{\end{eqnarray}}
\newcommand{\re}{{\rm Re\;}}
\newcommand{\im}{{\rm Im\;}}
\newcommand{\G}{{\cal G}}

\begin{document}
\title{Dynamical conductance in the two-channel Kondo regime of  a double dot system}
%\author{AUTHORS}
\author{A.\ I.\ T\'oth,$^{1,2}$ 
L.\ Borda,$^1$ 
J.\ von Delft$^3$  
and G.\ Zar\'and$^{1}$}
\affiliation{
$^1$ Theoretical Physics Department, Institute of Physics, 
Budapest University of Technology and Economics, H-1521 Budapest, Hungary
\\
$^2$ Institute for Theoretische Festk\"orper Physik, Universit\"at Karlsruhe,
  D-76128 Karlsruhe, Germany
\\
$^3$ Physics Department, Arnold Sommerfeld Center for Theoretical Physics and
Center for Nanoscience, Ludwig-Maximilians-Universit\"at, D-80333 Munich, Germany
}

\date{\today}
\begin{abstract}
We study finite-frequency transport properties of the 
double-dot system recently constructed to observe the two-channel Kondo 
effect [R.\ M.\ Potok {\em et al.}, Nature {\bf 446}, 167 (2007)]. We 
derive an analytical expression for the 
frequency-dependent linear conductance of this device in the Kondo regime. 
We show how the features characteristic of the 2-channel Kondo quantum 
critical point emerge in this quantity, which we compute using the results of
conformal field theory as well as  numerical renormalization group methods. 
We determine the  universal  cross-over functions describing 
non-Fermi liquid vs. Fermi liquid cross-overs  and also investigate the
effects of a finite magnetic field.  
\end{abstract}
\maketitle

\section{Introduction}

In the past few years semiconducting quantum dots have been in the focus of
intense research. This research  has 
mostly been motivated by their possible application in future
microelectronics: these devices behave as tunable artificial atoms
attached to electrodes,\cite{QD_review} 
they can be used as single electron transistors,\cite{SET} 
furthermore they also serve as a playground to model and study (artificial) molecular transport
in a very controlled way. They display a number of correlation-induced effects like 
the Coulomb blockade or the Kondo effect\cite{CB,Kondo} and they can also give rise to exotic
strongly correlated states.\cite{SU4} Very importantly, quantum dots can also be used
to build quantum bits with the electron spin providing the necessary degree
of freedom for quantum-computation.\cite{Loss} 

Nevertheless, maybe the most fascinating application of quantum dots is their possible 
use to realize  quantum phase  transitions between different correlated states. 
Several such transitions have been proposed: under special circumstances the
transition between the triplet and the singlet state of a dot can be a true
quantum phase transition,\cite{SchollerHofstetter,GoldhaberST} although in
most cases  this transition becomes  just a cross-over.\cite{WanderViel}
Dissipation can also lead to a quantum
phase transition where the  charge degrees of freedom of the dot become 
localized.\cite{LeHur,Simon} Unfortunately, these phase 
transitions have a Kosterlitz-Thouless structure and are -- in a sense -- 'trivial'
quantum phase transitions. Using multi-dot systems, however, it is also possible to 
realize {\em generic} quantum phase  transitions, where the transition point represents a true
critical state characterized by anomalous dimensions and a singular behavior. 
These critical states are generic non-Fermi liquid states in the sense that they cannot be
described in terms of conduction electron quasiparticles even at the Fermi 
energy.\cite{Nozieres}  The prototypes of these generic quantum impurity
states are the two-channel
Kondo model\cite{Cox} and the two-impurity Kondo model.\cite{2imp}  
Some years ago Matveev proposed that the two-channel Kondo model could be realized
by charge fluctuations at the charge-degeneracy point of a quantum dot.\cite{Matveev} 
However, Matveev's mapping 
assumes a vanishing level spacing and with present-day technology it has
been impossible to reach this state so far. However, a few years ago Oreg and
Goldhaber-Gordon proposed to realize the two-channel Kondo state 
through a double dot system,\cite{Oreg} and after several years of work this
two-channel Kondo state has
indeed been observed 
in a pioneering double dot experiment at Stanford.\cite{Potok}  For the 
realization of the other prototypical non-Fermi liquid state, the two-impurity
Kondo state,  a somewhat similar multi-dot setup has been proposed  recently.\cite{chung}

\begin{figure}[bp]
  \includegraphics[width=0.75\columnwidth,clip]{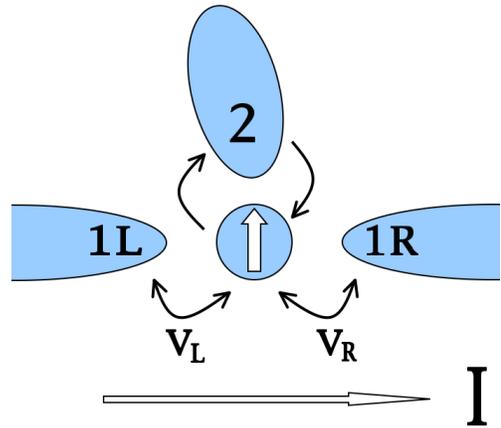}
%{2ck_setup.eps}
  \caption{Two-dot device: the small dot in the center couples to a large dot (2)
    and to a left and a right lead ($1L$ and $1R$) via the hopping amplitudes:
    $v_L$ and $v_R$. The small dot has a large level spacing, and the large dot is
    characterized by a vanishing level spacing, while both dots are in the
    Coulomb blockade regime. As a result, only spin exchange is possible between the dots.}
  \label{fig:2dot}
\end{figure}
Fig.~\ref{fig:2dot} shows the double dot device suggested by Oreg and Goldhaber-Gordon,
which has since been used 
to experimentally reach the two-channel Kondo fixed point \cite{Potok}.  
This set-up consist of a small dot coupled to a large dot (2) and two leads 
($1L$ and $1R$). The small dot is tuned to the regime
where charge fluctuations are suppressed and it has only one extra electron on
it.  The level spacing, $\delta\epsilon_s$, of the small dot and its charging
energy $\sim E_{Cs}$ are assumed to
be much larger than the temperature, $\delta\epsilon_s, E_{Cs} \gg T$, so that
below the scale $D$ charge fluctuations on the small dot are suppressed
and  the only role of this dot is to provide a spin. The size of the large dot, on the other hand,
is chosen in such a way that its charging energy and level spacing satisfy 
 $E_{C2} > T> \delta\epsilon_2$. This implies that this dot is also in the Coulomb
 blockade regime while the electronic states on it form a continuum of
 electron-hole excitations. Therefore, electrons on the large dot form a bath that 
can exchange spin with the small dot while electrons cannot jump out of
it\cite{Oreg}  as it is also indicated in Fig.~\ref{fig:2dot}. 
In the limit of small tunneling amplitudes, apart from some irrelevant and 
potential scattering terms, this double dot system is described by the 
following simple two-channel Kondo Hamiltonian, 
\begin{eqnarray}
H_{int}=\frac{1}{2}J_1{\vec S}\psi^{\dagger}_1{\vec
  \sigma}\psi^{}_1+\frac{1}{2}J_2{\vec S}\psi^{\dagger}_2{\vec \sigma}\psi^{}_2\;.
\label{eq:Kondo}
\end{eqnarray}
The operator $\psi_2$ describes electrons on the large dot.  In the
continuum limit, $\delta\epsilon_2 \to 0$, it  is defined as
\be
\psi_{2,\sigma} = \int a_\sigma(\epsilon) \;d\epsilon
\label{eq:psi2}
\ee
with  $a_\sigma(\epsilon) $ the annihilation operator of a conduction
electron of energy $\epsilon$ and spin $\sigma$ on the large dot, 
satisfying the anticommutation relation: 
$
\{a_{\sigma}(\epsilon),a^{\dagger}_{\sigma{'}}(\epsilon{'})\}=\delta_{\sigma\sigma{'}}\;\delta(\epsilon-\epsilon{'})
$.   
The operator  $\psi_1$ in Eq.~(\ref{eq:Kondo}) 
is a suitably chosen linear combination 
of electrons on the left and right lead-electrodes, 
\be
\psi_1=\frac{v_L\psi_L+v_R\psi_R}{(v_L^2+v_R^2)^{1/2}}
\ee
with $v_L$ and $v_R$ the hopping amplitudes between the dot and the 
left and right electrodes, respectively. The left and right 
fields $\psi_{L/R}$ are defined similarly to  Eq.~(\ref{eq:psi2}),
%G 
\be
\psi_{L/R,\sigma} = \int c_{L/R,\sigma}(\epsilon) \;d\epsilon\;,
\label{eq:psiLR}
\ee
with  $c_{L/R,\sigma}(\epsilon)$ the annihilation operator of a conduction
electron of energy $\epsilon$ and spin $\sigma$ on the left/right lead.

We remark that, strictly speaking, the Kondo Hamiltonian above 
is only accurate in the limit of small tunneling, while in the experiments
 the tunneling rates were quite large in order to boost up
the Kondo temperature.\cite{Potok}  Therefore, to study the region far 
above $T_K$, an Anderson model-type approach would be needed that also 
accounts for charge fluctuations of the small dot.\cite{Anders} 
Nevertheless, our Kondo model-based approach 
captures accurately the universal cross-over functions in the region of
interest, i.e.  around and far below the  Kondo temperature, 
provided that both dots are close to the middle of the Coulomb blockade 
regime. To account for deviations from the middle of the Coulomb blockade 
valley, one could break the particle-hole symmetry 
of Eq.~(\ref{eq:Kondo}) and add potential scattering terms to it. 

\begin{figure}[tp]
\begin{center}
  \includegraphics[width=.9\columnwidth,clip]{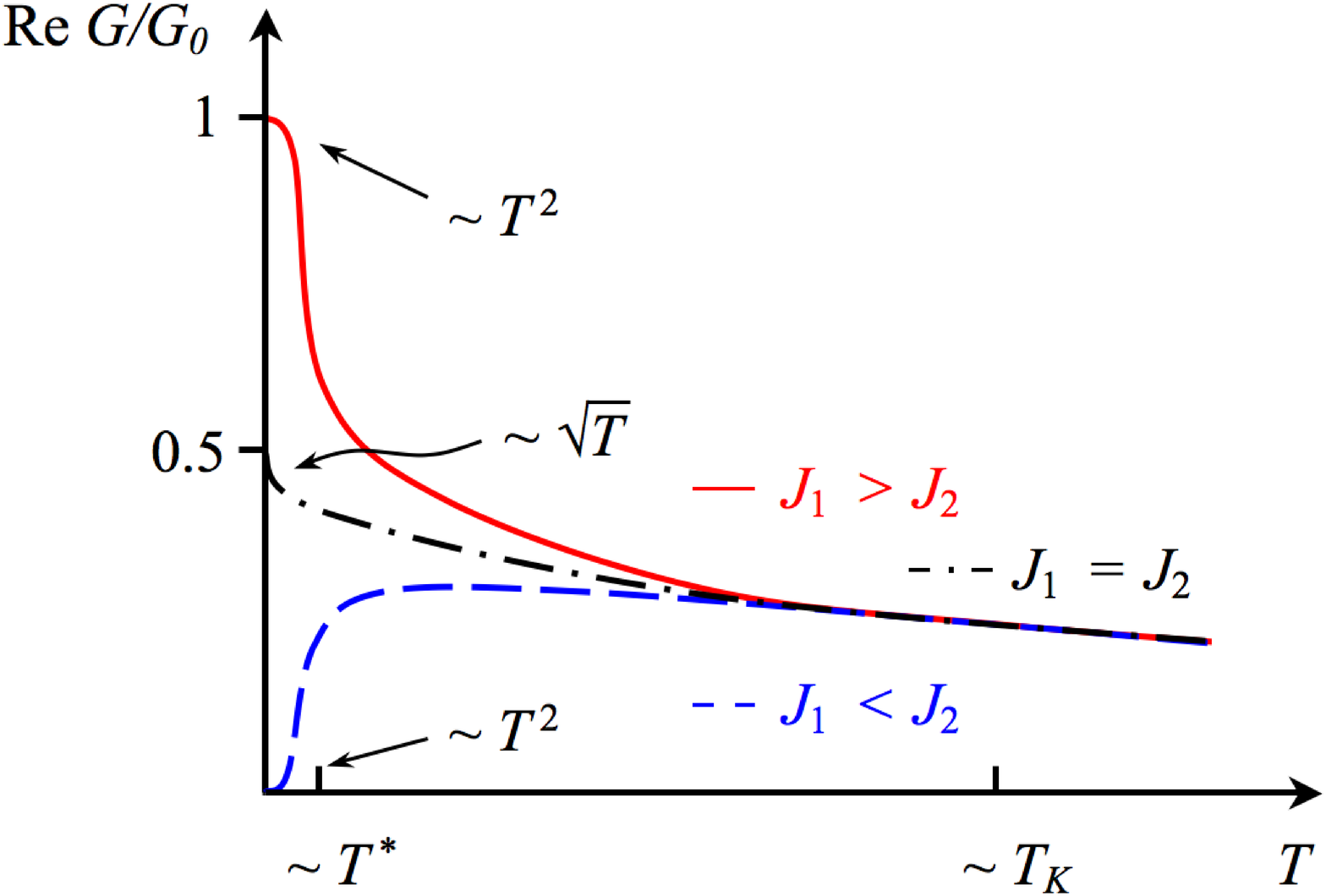}
\null
\vspace{1cm}
  \includegraphics[width=7cm]{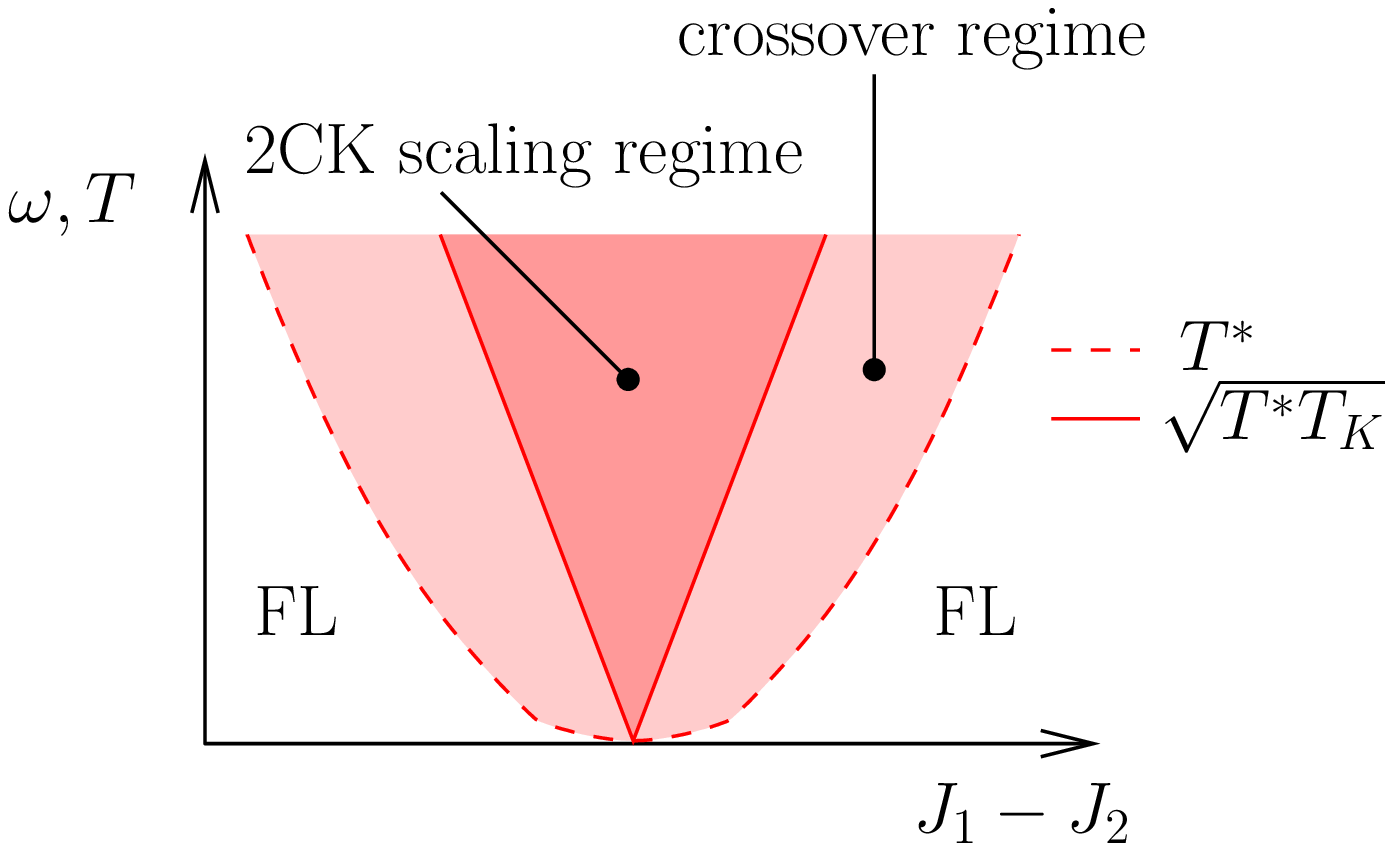}
\end{center}
%\label{fig:phasediag}
  \caption{Top: Sketch of the conductance through the small dot 
    divided by its maximum value, $G_0$, as a function of
    temperature. For $J_1=J_2$ a $\sim \sqrt{T}$ singularity emerges, while 
    for $J_1\ne J_2$ a Fermi liquid is formed at a scale $T^\ast$, and the
    conductance crosses over to a very small or a large value, with 
a characteristic 
    Fermi liquid scaling, $\sim (T/T^\ast)^2$. 
    Bottom: Sketch of the ``phase diagram'' of the two-channel Kondo model.
  }  
  \label{fig:conductance}
\end{figure}

The quantum critical state arises as a result of the competition 
of channels 1 and 2  to  form a singlet with the dot spin $S$.  
Depending on the values of the dimensionless couplings, $J_{1,2}$, 
two situations can occur:\cite{Oreg,Cox} 
(a) For  $J_1<J_2$ the spin  of the
small dot forms a Kondo singlet with electrons on the large dot 
that screen the spin at an energy scale $ T^\ast$. 
In this case, to promote a conduction electron between the left and right 
leads one needs to break up the Kondo singlet and pay 
an energy $T^\ast$,  and  
therefore  transport through the small dot is suppressed at low 
temperatures.\cite{Oreg,Potok} (b) For $J_1>J_2$, on the other hand,  
 the spin of the small dot is screened by  electrons in the leads. 
This correlated singlet state serves as a 'bridge' and
helps the lead electrons to propagate between the left and the right sides
with a small reflection probability and is thus characterized by a 
conductance of the order of the quantum conductance, $2e^2/h$. 
In both cases a Fermi liquid state is formed below the 
scale $T^\ast$, which can be characterized by simple phase shifts 
at the Fermi energy.\cite{Nozieres}

Interestingly, for  $J_1, J_2\to J$ the scale $T^\ast$
vanishes as  $T^\ast\sim (J_1-J_2)^2$,
and a non-Fermi liquid state emerges below the Kondo 
scale, $T_K\approx D \;e^{-1/J}$, with the cut-off $D$ defined as
$D\equiv \min\{\delta \epsilon_s, E_{Cs}, E_{C2}\}$.\cite{Cox}  
%($D=\min\{ \delta\epsilon_s,E_{C2}\}$)
This so-called two-channel Kondo state is characterized by a
conductance that is about half of the quantum conductance at very 
low temperatures, and has a $\sim\sqrt{T/T_K}$  
singularity for $T \ll T_K$.\cite{Oreg} 
This state  is -- in a sense -- a quantum
critical state:  although it is just 
a single point in the parameter space, 
it separates two stable Fermi-liquid phases, and
it influences the   behavior of the double dot system over the whole regime, 
$T^\ast < T,\omega< T_K$ for $J_1\approx J_2$. However, as we shall see later, 
the scaling properties usually associated with  the two-channel Kondo fixed
point itself are  restricted to a somewhat smaller energy range, 
$\sqrt{T^\ast T_K} < T,\omega< T_K$. The characteristic features of the
temperature-dependence of the DC conductance  and the schematic phase diagram
are sketched in  Fig.~\ref{fig:conductance}. 

The purpose of the present paper is to investigate {\em dynamical} transport
properties of the above set-up and determine how the two-channel Kondo
behavior and the presence of a quantum critical point at $J_1=J_2$ 
manifests itself in the 
AC conductance through the dot. For this purpose, we shall derive 
an expression for the AC conductance in the linear response regime that
relates the conductance to the so-called composite Fermions' propagator at
any temperature and frequency. Making use of  this simple formula, we shall
evaluate the AC conductance for frequencies $T\ll \omega$ using numerical
renormalization group methods. We shall also determine the universal
cross-over functions that describe the non-Fermi liquid vs. Fermi liquid 
cross-over for $T^\ast \ll T_K$. 
As we show, the AC conductance exhibits features that are qualitatively 
similar to the finite temperature DC conductance, 
sketched in Fig.~\ref{fig:conductance}. In addition, we shall also 
investigate, what conclusions we can draw regarding AC properties 
based upon the predictions of conformal field theory, and use this approach 
to obtain the universal scaling of the conductance in the regime 
$T^* \ll \omega,T \ll T_K$ .

The paper is organized as follows. Section \ref{sect2}  provides the details
of the derivation of the AC conductance formula for the two-channel Kondo
model. In Section \ref{sect3} we 
present some  analytical considerations 
based on conformal field theory  
concerning the universal scaling 
properties  of the linear conductance and of the eigenvalue of 
the so-called on-shell 
$T$-matrix. Section \ref{sect4} comprises our numerical renormalization 
group results for the composite Fermions' spectral function and the linear
conductance in case of channel anisotropy and in the presence of a magnetic 
field. At last our conclusions are summarized. 

\section{Kubo formula and composite Fermions\label{sect2}}

Let us start our analysis with
the derivation of a simple expression for the AC conductance in 
terms of the so-called composite Fermion operators.\cite{Theo}
For this purpose, we 
first couple an external voltage to the dot and introduce a 
time-dependent chemical potential difference between the left and right
electrodes: 
\be
H_V \equiv V(t) \;Q = e\;V(t) \;( N_R-N_L)\;,
\ee
with $N_R$ and $N_L$ the number of electrons in the left and right leads,
respectively, 
$$
N_{L/R}= \sum_{\sigma}\int
c_{L\sigma}^{\dagger}(\epsilon)c^{}_{L\sigma}(\epsilon) \;d\epsilon\;.
$$ 
The current operator can be defined as the time derivative of $Q$, 
$I(t) = i \;[H,Q(t)] =i \;[H_{\rm int},Q(t)]$. This commutator is easily 
shown to give
\be
I=e \; 
%\alpha_L \alpha_R
\frac{v_L v_R}{v_L^2 +v_R^2}\;
J_1\left(iF^{\dagger}_1{\tilde\psi^{}_1}+h.c.\right)\;,
\ee
where $\tilde \psi_1$ denotes the decoupled electron field of the 
leads,
\be
{\tilde \psi}_1=\frac{v_L\psi_L-v_R\psi_R}{(v_L^2+v_R^2)^{1/2}}\;,
\ee 
and we have introduced the so-called composite Fermion operator,
\be 
F_\sigma^{\dagger}\equiv
\sum_{\sigma'}\psi_{1\sigma'}^{\dagger}{\vec\sigma}_{\sigma'\sigma }{\vec S}\;.
\ee
The operator $F^{\dagger}$ has spin 1/2 and charge 1, and it corresponds to the
'universal part' of  the electron localized on the small dot. 

Close to equilibrium, the current  through the dot is 
given by the Kubo formula
\bea
\langle  I(t)\rangle &=& \int G(t-t{'})\; V(t')\;dt'\;,
\nonumber
\\ 
G(t-t{'})&=&i\left<\left[I(t),Q(t{'})\right]\right>\theta(t-t{'}),
\eea
with $G(t-t{'})$ the conductance.
Differentiating with respect to time and then taking the Fourier 
%transformed
transform  we obtain the relation 
\be
-i\omega \;G(\omega)=\G^R_{II}(\omega)-A,
\ee
where $\G^R_{II}$ denotes the retarded current-current correlation function
and $A$ is a real constant
\be
A = i \left< [Q(t'),I(t')]\right> = \G^R_{II}(\omega=0)\;.
%A=-4J_1 \frac{t^2_Rt^2_L}{t^2_R+t^2_L} \left<{\vec
%  S}\psi^{\dagger}_1{\vec \sigma}\psi_1\right> \;.
\ee
Thus the real and imaginary  parts of the conductance are given by 
\bea
{\rm Re}\{G(\omega)\}&=&- \frac 1 \omega {\rm Im}\{\G^R_{II}(\omega)\}\;,
\\
{\rm Im}\{G(\omega)\}&=& \frac 1 \omega \left(
{\rm Re}\{\G^R_{II}(\omega)\} - 
{\rm Re}\{\G^R_{II}(0)\} 
\right) 
\;.
\eea
In general, it is not so simple to compute the correlation function
$\G^R_{II}$. In our case, however, the field  $\tilde \psi_1$ 
is completely decoupled from the spin and describes non-interacting 
Fermions. This observation allows us to write  $\G^R_{II}(t)$ as
\begin{widetext}
\bea
\G^R_{II}(t) = -i\; e^2\;\frac {v_R^2 v_L^2}{(v_R^2 + v_L^2)^2} J_1^2 \;
\sum_\sigma \Bigl[\G^R_{F\sigma}(t) \G^<_{\tilde\psi\sigma}(-t) +
\G^<_{F\sigma}(t) \G^A_{\tilde\psi\sigma}(-t)
+  \G^R_{\tilde\psi\sigma}(t) \G^>_{F\sigma}(-t)+  \G^>_{\tilde\psi\sigma}(t) \G^A_{F\sigma}(-t)
\Bigr]\;
\label{eq:G_II_realtime}
\eea
where $\G^R$,  $\G^A$,  $\G^>$, and $\G^<$ denote the usual retarded, advanced,
bigger and lesser Keldysh Green's functions. The Fourier  transform of this
expression simplifies considerably if one uses the fact that the field   
$\tilde \psi_1$ is non-interacting and therefore the corresponding Green's
functions become in the large band-width limit
\bea
\G^R_{\tilde\psi\sigma}(\omega) = \G^A_{\tilde\psi\sigma}(\omega)^\ast =
-\frac i 2\;,\phantom{nnn}
\G^<_{\tilde\psi\sigma}(\omega) = i \; f(\omega)\;,
\eea
with $f(\omega)$ the Fermi function. Taking the real and imaginary parts of the Fourier
%transformed 
transform of Eq.~\eqref{eq:G_II_realtime}  we finally obtain:
\bea 
{\rm Re\;}\{G(\omega)\} &=& 
\frac {G_0}{8\;\omega}
%\frac {2 e^2}h \sum_\sigma \frac{ 4 v_L^2 v_R^2}{(v_L^2 + v_R^2)^2}
\sum_\sigma
\int {d\omega'} \;
{\rm Im}\;\{t_\sigma(\omega')\}\;
\bigl[f(\omega'+\omega) - f(\omega'-\omega)\bigr]\;, 
\label{eq:ReG_tmatrix}
\\
{\rm Im\;}\{G(\omega)\} &=& 
\frac {G_0}{8\;\omega}
%\frac {2 e^2}h \sum_\sigma \frac{ 4 v_L^2 v_R^2}{(v_L^2 + v_R^2)^2}
\sum_\sigma \int 
{d\omega'} \;
{\rm Re}\;\{t_\sigma(\omega')\}\;
\bigl[f(\omega'+\omega) + f(\omega'-\omega)-2 f(\omega')\bigr]\;, 
\label{eq:ImG_tmatrix}
\eea
\end{widetext} 
where we introduced the dimensionless eigenvalue 
$t_\sigma(\omega)$ of the  so-called on-shell  $T$-matrix,\cite{dephasing_2} 
which describes the scattering of electrons of energy $\omega$,  
\be
t(\omega)= -J_1^2\; \G^R_{F\sigma}(\omega)\;,
\ee
and $G_0$ denotes the maximum conductance through the dot, 
\be
G_0= \frac {2 e^2}h  \frac{ 4 v_L^2 v_R^2}{(v_L^2 + v_R^2)^2}\;.
\ee
Thus the real part of the conductance is related to the imaginary  part of
$\G^R_{F\sigma}$, which  is essentially  the spectral function 
of the composite Fermion,  $\varrho_{F\sigma}(\omega)$.  The latter can be
determined numerically using the numerical renormalization group method. 
Then the real part, $ {\rm Re}\{G^R_{F\sigma}\}$, can  be obtained by performing a
 Hilbert transformation numerically, and the imaginary part of the conductance 
can then be calculated from  $ {\rm  Re}\{G^R_{F\sigma}\}$ by simple numerical integration. 
Note that 
Eqs.~(\ref{eq:ReG_tmatrix}) and (\ref{eq:ImG_tmatrix}) provide the linear conductance  through the dot 
for any asymmetry parameter at any temperature and any frequency. They are thus 
natural extensions of the formula given in
Ref.~\onlinecite{PustilnikGlazman?}, and are the analogues 
% the analogy of sg = valami analogiaja
% the analogue of sg = valami analogja
of the formulas obtained 
recently for the Anderson model.\cite{sindel05}
%
%
%\bea 
%{\rm Re\;}\{G(\omega)\} &=& 
%\frac {e^2}h\; \frac {J_1^2} \omega \sum_\sigma \frac{ v_L^2 v_R^2}{(v_L^2 + v_R^2)^2} 
%\int 
%{d\omega'} \; {\rm Im}\{\G^R_{F\sigma}(\omega')\}\;  
%\bigl[f(\omega'+\omega) - f(\omega'-\omega)\bigr]\;, 
%\label{eq:ReG}
%\\
%{\rm Im\;}\{G(\omega)\} &=& 
%\frac {e^2}h\; \frac {J_1^2} \omega 
% \sum_\sigma \frac{ v_L^2 v_R^2}{(v_L^2 + v_R^2)^2} 
%\int 
%{d\omega'} \; {\rm Re}\{\G^R_{F\sigma}(\omega')\}\;  
%\bigl[2f(\omega') - f(\omega'+\omega) - f(\omega'-\omega)\bigr]\;. 
%\label{eq:ImG}
%\eea
%
%

Eqs.~(\ref{eq:ReG_tmatrix}) and (\ref{eq:ImG_tmatrix})
belong to  the main results of our paper. 
We shall use these formulas to compute the AC conductance through the dot in the vicinity of
the two-channel Kondo fixed point. 

\section{Analytical considerations\label{sect3}}

Eq.~(\ref{eq:ReG_tmatrix}) allows us to make numerous statements based on
rather general properties of the two-channel Kondo fixed point.\cite{Cox}
From an
exact theorem of Affleck and Ludwig,\cite{AffleckLudwig} e.g. we know that at the two-channel
Kondo fixed point (i.e. for $J_1=J_2$ and $\omega,T\to 0$) the $S$-matrix of
the conduction electrons identically vanishes.  From the relation,
$S(\omega) = 1 + i \; t(\omega)$  between the dimensionless eigenvalue of the
$S$-matrix and the $T$-matrix we thus obtain
\be 
\lim_{\omega,T\to0}\; t(\omega,T)= i \phantom{nnnn}(J_1=J_2)\;.
\label{eq:t=i}
\ee
From this, it immediately
follows that  at the two-channel Kondo fixed point the conductance takes half
of its maximum value,
\be
\lim_{\omega,T\to0}\;G(\omega,T)= G_0/2\;,\phantom{nnn}(J_1=J_2)\;.
\ee
The results of conformal  field theory\cite{AffleckLudwig}
  also enable us to compute 
the finite frequency conductance for  $J_1=J_2$ and 
$\omega,T\ll T_K$. In this limit the $T$-matrix is given by the 
expression\cite{AffleckLudwig} 
\begin{widetext}
\bea
t(\omega)={i}\left\{1-3\left(\pi\; T \right)^{1/2}\lambda\int\limits_0^1
du\left[
u^{-i\beta\omega/2\pi}u^{-1/2}(1-u)^{1/2}F(u)-\frac{4}{\pi}u^{-1/2}(1-u)^{-3/2}\right]
\right\}\;,
\label{eq:CFT_Tmatrix}
\eea
\end{widetext}
where $F(u)\equiv F(3/2,3/2,1;u)$ is the hypergeometric function,
 and  $\lambda$ stands for the amplitude of the leading
irrelevant operator:
\be
\lambda = \frac {\gamma}{\sqrt{T_K}}\;. 
\ee 
The  value of the dimensionless constant $\gamma$ depends on the precise 
definition of $T_K$. Throughout this paper, we shall define $T_K$ as the
energy at which for $J_1=J_2$  the composite Fermion's spectral function 
drops to half of its value, ${\rm Im}\;t(\omega=T_K) = {\rm Im}\;t(\omega=0)/2$. 
Then, comparing the numerical results of Section~\ref{sect4} to the asymptotic
$\omega\gg T$ behavior of the conductance we obtain the value
$\gamma = 0.093\pm0.001$. Clearly, since the omega-dependence 
enters $t(\omega)$ only in the combination $\omega/T$, it immediately follows that
$1-\im t(\omega,T)/(\lambda T^{1/2})$ 
is a universal function of $\omega/T$ (see inset of Fig. 3). 

\begin{figure}[bp]
\includegraphics[width=0.9\columnwidth,clip]{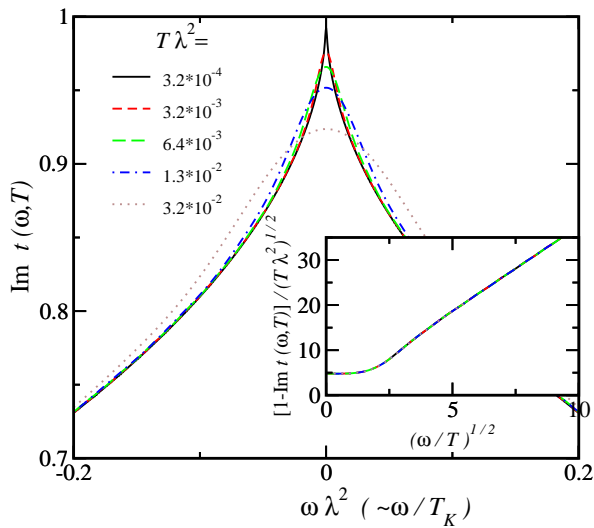}
\caption{Imaginary part of the eigenvalue of the T-matrix  obtained 
by numerical integration of Eq.\eqref{eq:CFT_Tmatrix}. 
The scale of the $\omega$ axis is set by the amplitude of the leading irrelevant
operator, $\lambda$. The inset illustrates how the curves corresponding to different
temperatures collapse into one universal curve. 
}
\label{fig:Im_CFT_Tmatrix}
\end{figure}

In Fig.~\ref{fig:Im_CFT_Tmatrix} we show the results obtained by numerically integrating
 Eq.\eqref{eq:CFT_Tmatrix} for a few  temperatures.
It is remarkable that  curves corresponding to different
temperatures cross each-other. 
This feature is a direct consequence of the unusual shape of the universal
curve shown in the inset of Fig.~\ref{fig:Im_CFT_Tmatrix}. 

\begin{figure}[bp]
\includegraphics[width=0.9\columnwidth,clip]{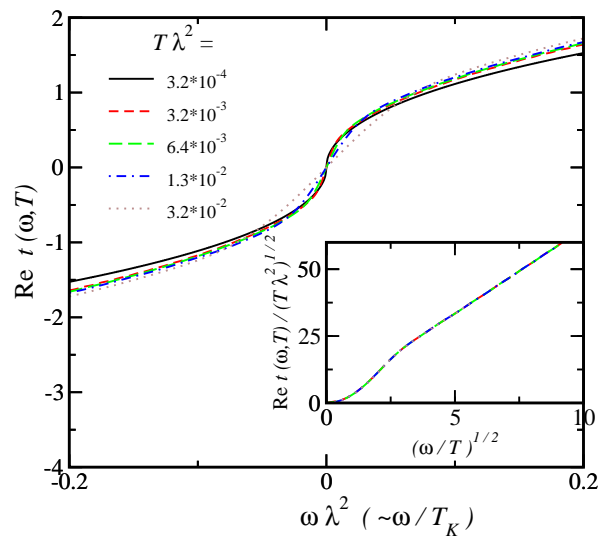}
\caption{Real part of the eigenvalue $t(\omega)$ of the T-matrix
predicted by conformal field theory. The inset shows 
the collapse to a single scaling curve (obvious from the integral definition). 
}
\label{fig:Re_CFT_Tmatrix}
\end{figure}

Note that to construct the universal scaling curve one needs to rescale the
axes with respect to the temperature only, and  the precise value of the Kondo temperature
appears only through the prefactor $\lambda$. The fact that the only relevant energy scale is the
temperature is characteristic of quantum critical points. The imaginary part
of the $T$-matrix exhibits  a 
$\sim\sqrt{|\omega|}$ cusp for $T \ll \omega\ll T_K$, and crosses
over to a quadratic regime for $\omega\ll T$. Similar behavior is observed in
the real  part of $t(\omega)$, shown in
Fig.~\ref{fig:Re_CFT_Tmatrix}. This quantity also shows a characteristic 
$\sim\sqrt{\omega}$ behavior at frequencies $T_K\gg \omega\gg T$, that crosses
over to a linear regime for $\omega\ll T$.

\begin{figure}[bp]
\includegraphics[width=0.9\columnwidth,clip]{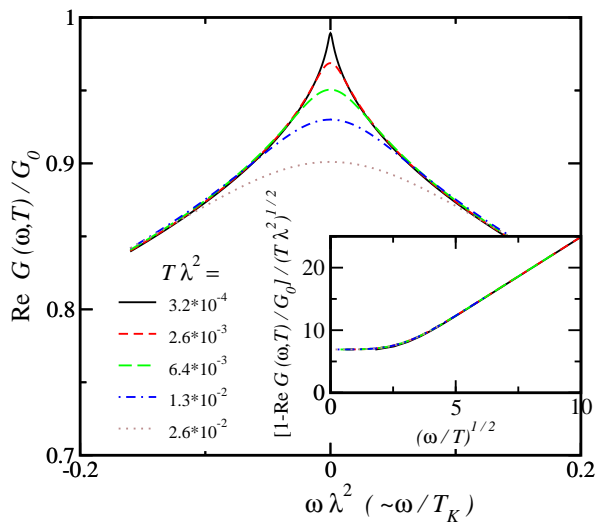}
\caption{Real part of the conductance computed
from  Eqs.\eqref{eq:CFT_Tmatrix},  
Eqs.\eqref{eq:ReG_tmatrix}, and \eqref{eq:ImG_tmatrix}. The inset shows the
universal collapse. 
}
\label{fig:ReG_CFT_new}
\end{figure}

\begin{figure}[tp]
\includegraphics[width=0.9\columnwidth,clip]{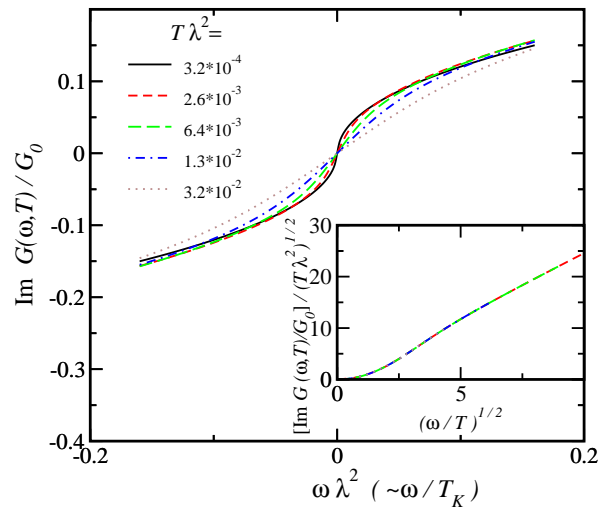}
\caption{Imaginary part of the conductance 
from  Eqs.\eqref{eq:CFT_Tmatrix},  
Eqs.\eqref{eq:ReG_tmatrix}, and \eqref{eq:ImG_tmatrix}. The inset shows the
universal scaling curve. 
}
\label{fig:ImG_CFT_new}
\end{figure}

Using Eqs.\eqref{eq:CFT_Tmatrix},  Eqs.\eqref{eq:ReG_tmatrix} and
\eqref{eq:ImG_tmatrix}, 
both the real and the imaginary parts of the 
conductance can be computed by numerical integration. 
The results are plotted in Figs.~\ref{fig:ReG_CFT_new} and \ref{fig:ImG_CFT_new}
for various temperatures. Even though, at first sight, the
results for the conductivity look qualitatively similar to those for the
$T$-matrix, there is an important difference: integration with the Fermi functions
apparently eliminated the aforementioned crossing of the curves. 
Similar scaling curves have been computed using conformal field theory results 
for the differential conductance of two-channel Kondo scatterers in point
contacts.\cite{vonDelftLudwigAmbegaokar}

In the limit, $T_K\gg \omega\gg T$, the conformal field theory  also predicts that 
the AC conductance scales  as
\bea 
%\re G (T_K\gg\omega\gg T) 
\re G (\omega) 
&\approx& \frac {G_0} 2 \Bigl(1-\alpha \sqrt{\omega\over T_K}\Bigr)\;,
\nonumber
\\
\im G (\omega) &\approx& \frac {G_0} 2 \;\alpha  \;{\rm sign}(\omega)
\;\sqrt{|\omega|\over T_K}\;,
\label{eq:G(om)_anal}
\eea
with $\alpha=2.53 \pm0.06$ a universal constant of order of unity. 
The fact that the coefficients in the  two equations above are both equal to $\alpha$,  
follows from the observation that $G (\omega)$ is analytical in the 
upper half-plane.

For $J_1\ne J_2$ a new Fermi liquid scale, $T^\ast$ (mentioned earlier),
emerges, but one can still make many statements based upon the fact 
that the  leading relevant and irrelevant operators  have scaling dimensions
$y_+=1/2$ and $y_-=-1/2$, respectively.\cite{Cox} 
As a consequence,  in the vicinity of the two-channel Kondo fixed point  
($T^\ast \ll T_K $) the conductance becomes a function of the form:
\be 
G(\omega,T) = G\Bigl(\frac \omega{T_K},\frac T {T_K},\frac{T^\ast}{ T_K}
\Bigr)\;,
\ee
with the Fermi liquid scale $T^\ast$ approximately given by 
\be 
T^\ast \approx T_K\;K_R^2\; \sim \;(J_1-J_2)^2\;,
\label{eq:Tstar}
\ee
where we introduced  the renormalized anisotropy parameter $K_R$ 
as 
\be  
K_R \equiv   \frac{4\left(J_1-J_2\right)}{\left(J_1+J_2\right)^2}\;.
\ee
Throughout this paper we shall define $T^\ast$ as the energy 
scale at which $\im t(\omega=T^\ast)= 1.5$ in the channel of larger coupling. 
Note that the parameter  $K_R$ can be considerably larger than 
the naive estimate, 
$(J_1-J_2)/(J_1+J_2)$ due to the   renormalization of the couplings 
$J_1$ and $J_2$ in the high energy regime, $D>\omega > T_K$.
In  the limit of $T^\ast, \omega \ll T_K$ and $T\to 0$ the conductance 
$G(\omega,T)$ becomes a universal function of $\omega/T^\ast$, 
\be
 G_{\{\omega,T^\ast\}\ll{ T_K} }\bigl( \omega,T=0 \bigr)= G_0\;F_{\pm}(\omega/T^\ast)\;.
\ee
The signs $\pm$ refer to the cases $J_1>J_2$ and $J_1<J_2$,
respectively, and 
 the scaling functions $F_\pm(y)$ have the properties
\be
\re F_\pm\left(\frac{\omega}{T^\ast}\right) \approx \left 
\{ 
\begin{tabular}{ll}
 $a_\pm + b_\pm \;   \left({\omega\over T^\ast}\right)^2$, & \phantom{nn}${\omega\ll T^\ast}$\;, \\
$1/2 \pm  c  \;\left({T^\ast\over \omega}\right)^{1/2} $, & \phantom{nn} ${\omega\gg T^\ast}$\;.
\end{tabular}
\right. 
\label{eq:cross-over}
\ee
In other words, for $\omega\ll T^\ast$ the conductance through the dot 
is Fermi liquid like, and $\re G$ shows a $\sim (\omega/T^\ast)^2$
behavior, while for $T_K\gg \omega\gg T^\ast$ the real part of the conductance scales to its 
two-channel Kondo value with a small but increasing correction,  $\sim
\sqrt{T^\ast/\omega}$. The latter behavior breaks down once the 
amplitude of the leading irrelevant operator, $\sim \sqrt{\omega/T_K}$, reaches 
that of the anisotropy operator, $\sim \sqrt{T^\ast/\omega}$, i.e. 
at frequencies in the range $\omega\approx \sqrt{T_K T^\ast}$.
The constants $a_\pm$, $b_\pm$, and $c$ above are numbers of  order unity that
depend somewhat on electron-hole symmetry breaking, but close to electron-hole
symmetry $a_+\approx 1$, and $a_-\approx 0$.  Note that the precise value of 
the constants $b_\pm$ and $c$  depends also on the definition of the scale 
$T^\ast$.

The imaginary part of  $F_\pm(y)$ has somewhat different properties and 
 behaves as  
\be
%\im F(y)_\pm \approx \left 
%\{ 
%\begin{tabular}{ll}
% $d_\pm    y$ & $y\ll 1$ \\
%$ e_\pm/\sqrt{y}$ & $y\gg 1$
%\end{tabular}
\im F_\pm\left({\omega\over T^\ast}\right) \approx \left 
\{ 
\begin{tabular}{ll}
 $ d_\pm \;   {\omega\over T^\ast}$, & for ${\omega\ll T^\ast}$\;, \\
$\pm  e  \;\left({T^\ast\over \omega}\right)^{1/2} $, & for ${\omega\gg T^\ast}$\;.
\end{tabular}
\right.% \;.
\label{eq:im_cross-over}
\ee
In other words, the imaginary part of $G$ must show a bump of size $\sim G_0$ at frequencies
$\omega\sim T^\ast$. These expectations shall indeed be met by our numerical results.

Similar to channel asymmetry, an external magnetic field also 
suppresses the non-Fermi liquid behavior,\cite{Cox} and introduces a new Fermi
liquid scale,
\be 
T_B \equiv \frac {B^2}{T_K}\;.
\ee
However, the magnetic field does not result in such a dramatic change in the
conductance as the channel-symmetry breaking: while at $\omega=0$ the
conductance exhibits a {\em jump } as a function of the channel-anisotropy,
it  changes continuously 
as a function of the magnetic field and shows only a cusp,\cite{LaciGlazman,Anders}
\be
G(B)_{J_1=J_2} \approx \frac {G_0} 2 \Bigl(1-\beta  \;\frac {|B|}{T_K} \;\ln( T_K/|B|)\Bigr)\;,
\ee
as it obviously follows from the singular behavior of the conduction electron 
phase shifts at the Fermi energy.\cite{AffleckLudwigPangCox,LaciGlazman} 
As we shall see later, the AC conductance displays much more interesting  
features in a  finite magnetic field.

\section{Numerical results\label{sect4}}

\begin{figure}[pb]
  \includegraphics[width=0.9\columnwidth,clip]{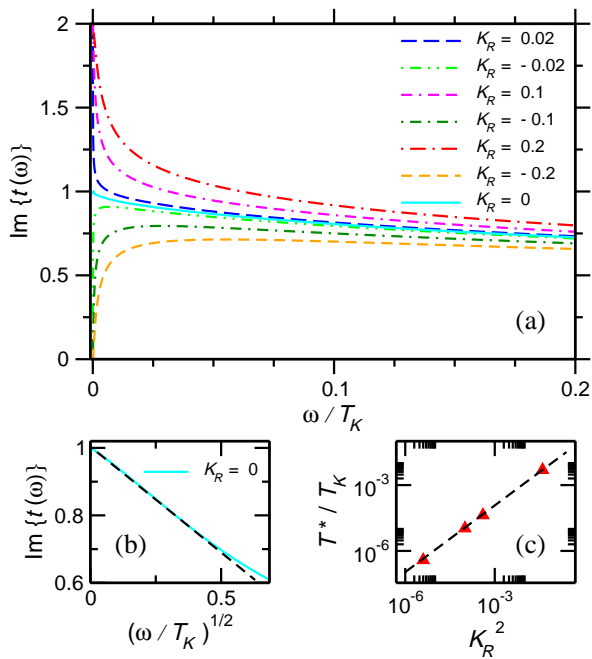}
  \caption{
(a) Imaginary part of the eigenvalue of  the on-shell T-matrix, as function of
$\omega/T_K$,  for several different values of the anisotropy parameter, $K_R= 4 (J_1 - J_2)/(J_1 + J_2)^2$. 
In all cases     $J_1 + J_2 = 0.2$. Curves with  $J_1>J_2$ or $J_1<J_2$ scale to 
$\im t(0) = 2$  or $\im t(0) = 0$, respectively.
    The critical curve corresponding to $J_1=J_2$ separates these two 
    sets of curves. 
(b) $\im t(\omega)$ for  $J_1=J_2$, as a function of 
    $\sqrt{\omega/T_K}$. The dashed line is a guide to the eye.
(c) $T^*$ as the function of $K_R^2$.
}
\label{fig:T}
\end{figure}

In this section, we shall use the numerical renormalization group (NRG) 
method\cite{NRG_ref}
to compute the eigenvalue of the $T$-matrix
and from that the AC conductance. Although Eqs.~(\ref{eq:ReG_tmatrix}) and 
(\ref{eq:ImG_tmatrix}) hold at any temperature, 
finite temperature calculations  are extremely delicate close 
to a non-Fermi liquid state.  Therefore, we shall present numerical
results  only for $T=0$ temperature here. 
Nevertheless, according to the basic principles of scaling, a finite 
frequency $\omega$ plays a role rather similar to that of a finite temperature, 
and therefore the $T=0$ temperature AC conductance, $G(\omega,T=0)$,
behaves rather similarly to the DC conductance at a finite temperature $T$, $G(\omega=0,T)$.

To perform accurate calculations we assumed an electron-hole symmetrical
conduction band and strongly exploited the symmetries of the Hamiltonian.
The numerical results presented here have been 
obtained using a new ``flexible'' NRG code, that handles symmetries 
dynamically.\cite{Anna} 
In particular, in the absence of an external magnetic field, we used a symmetry
$SU_{c1}(2)\otimes SU_{c2}(2)\otimes SU_s(2)$, 
with  $SU_{c1}(2)$ and $SU_{c2}(2)$ the charge
$SU(2)$ symmetries in channels 1 and 2, respectively,\cite{Jones} 
and  $SU_s(2)$ the spin
$SU(2)$ symmetry.  
The advantage of this symmetry is that it is not violated
even for $J_1\ne J_2$, and it breaks down only to  $SU_{c1}(2)\otimes
SU_{c2}(2)\otimes U_s(1)$ in the presence of a magnetic field. For the channel
anisotropic cases we have retained a maximum of 750 multiplets during the NRG
calculations, whereas 850 multiplets were kept in the presence of a magnetic 
field. All calculations were carried out with a discretization parameter $\Lambda=2$.
To compute the AC conductance, we have determined the 
composite Fermion's spectral function which, apart from an overall
normalization factor, is equal to $\im t(\omega)$. 
This normalization factor can be easily fixed for $J_1=J_2$ 
using the condition, Eq.~(\ref{eq:t=i}). This procedure is 
much more accurate than estimating the normalization factor from the bare
couplings, since the latter procedure suffers from the NRG discretization 
problem as well as from the loss of spectral weight at high energies, leading 
generally to a few percent error in the amplitude. 

\subsection{Channel symmetry breaking}
First we investigated numerically how the non-Fermi liquid 
structure appears in the AC conductance through the 
double dot and how channel anisotropy destroys this non-Fermi 
liquid behavior.
Some typical results are shown in Fig.~\ref{fig:T}:  for $J_1 = J_2$ 
we recover the two-channel Kondo result,  $\im t(\omega\to0)=1$, and 
the deviation from the fixed point value scales as $\sim\sqrt{\omega/T_K}$, in
agreement with Eq.~(\ref{eq:G(om)_anal}).

\begin{figure}[bt]
\includegraphics[width=0.9\columnwidth,clip]{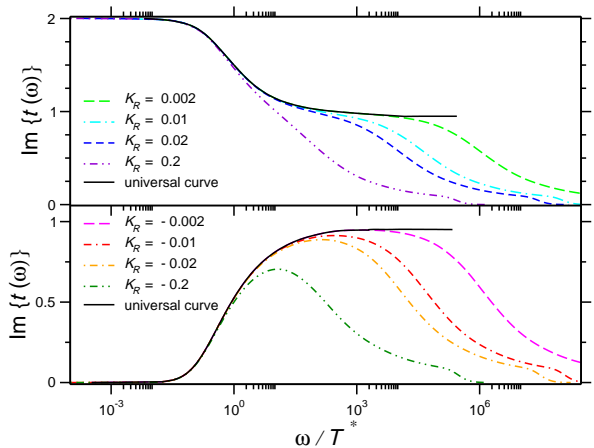}
\caption{Imaginary part of the on-shell $T$-matrix in the presence of channel
  anisotropy as the function of  $\omega/T^*$.
  The upper part corresponds to $J_1>J_2$ while the
  lower part to $J_1<J_2$. In both cases for $T^\ast,\omega \ll T_K$ the
  curves 
  %below the frequency $T^*$
  follow the universal cross-over  function, corresponding to a 
  $(\omega/T^\ast)^2$-like scaling
  at low frequencies and a $1 \pm  c \; (T^\ast/\omega)^{1/2}$ behavior at
  large frequencies.   
} 
\label{fig:T_scaled_doubled}
\end{figure}

For $J_1 \ne J_2$ the new cross-over scale $T^\ast$ appears below which 
$\im t(\omega)$ crosses over from the two-channel Kondo value $\im
t(\omega)= 1$, to $\im
t(\omega)= 2$ for $J_1 > J_2$ or to $\im t(\omega)= 0$ for $J_1 <
J_2$ in the electron-hole symmetrical situation studied numerically.  
In the limit $T^\ast\ll T_K$ this cross-over is described by   universal
cross-over functions, similar to Eq.~(\ref{eq:cross-over}). 
We determined these scaling functions  numerically and displayed them in
Fig.~\ref{fig:T_scaled_doubled}. (The black curves were obtained by taking an
extremely small value of $T^\ast$, and chopping off the parts near $\omega\sim T_K$.)
The Fermi liquid  scale $T^\ast$ extracted
from $t(\omega)$ is  shown in Fig.~\ref{fig:T}.(c), and is in excellent agreement with the analytical  
expression, Eq.~(\ref{eq:Tstar}).

\begin{figure}[tb]
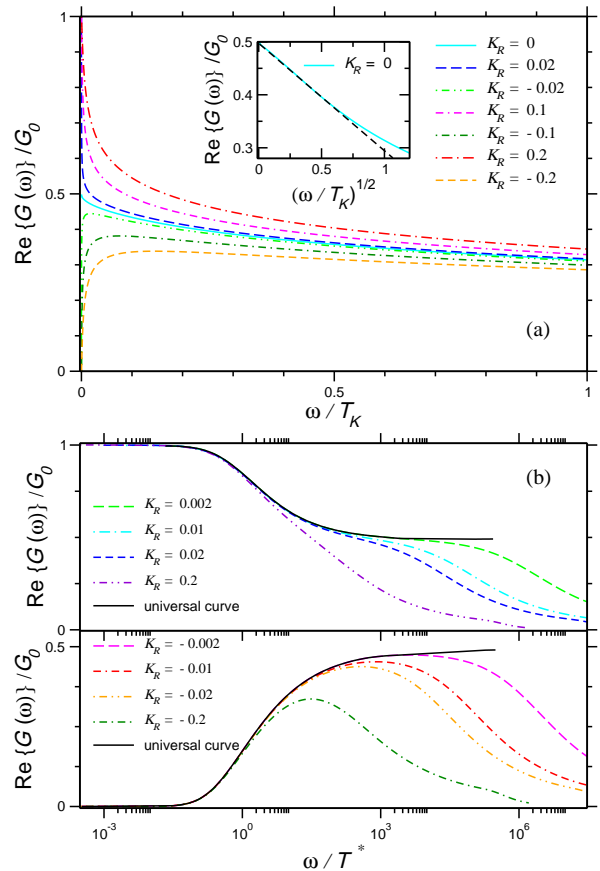

  \includegraphics[width=0.9\columnwidth,clip]{G_chani.eps}
  \includegraphics[width=0.9\columnwidth,clip]{G_chani_Tstar.eps} 
  \caption{(a) AC conductance as the function of $\omega/T_K$. 
    For   $J_1>J_2$  and $J_1<J_2$ the curves  scale 
    $\re G \to G_0$ and $\re G \to  0$, respectively.
    Inset: AC conductance for $J_1=J_2$  as the function of $\sqrt{\omega/T_K}$. 
    (b) AC conductance for positive (upper part) and negative (lower part)
    channel anisotropy parameters as the function of $\omega/T^*$.
For $\omega, T^\ast\ll T_K$, 
    the curves follow the universal cross-over curves.
}
\label{fig:G}
\end{figure}

According to Eq.~(\ref{eq:ReG_tmatrix}), the  real part of 
the conductance can be computed  from $\im t(\omega)$ through a 
simple integration. The resulting conductance curves are shown in Fig.~\ref{fig:G}.
The behavior of $\re G(\omega)$ is strikingly 
similar to that of $\im t$: it also exhibits a 
$\sim \sqrt{\omega}$ singularity for $J_1=J_2$ and 
crosses over from a value $G= G_0/2 $
to  $G=  G_0$ or to  $G= 0$ at the scale 
$T^\ast$ following the universal cross-over functions, 
$F_\pm(\omega/T^\ast)$. We remark here that there seems to be no 
other reliable way than NRG to determine these universal 
cross-over functions, which connect two separate strong coupling fixed points, 
the non-Fermi liquid  liquid fixed point and the Fermi liquid fixed point. 
These universal cross-over functions constitute some of the central
results of this work.

\begin{figure}[tb]
\includegraphics[width=0.9\columnwidth,clip]{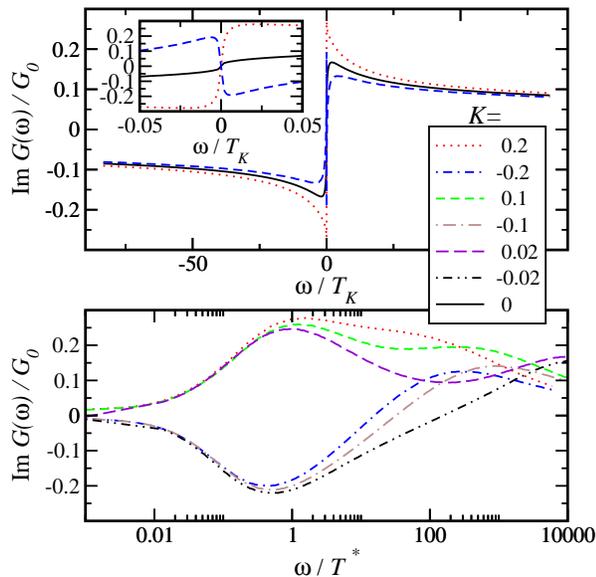}
\caption{Imaginary part of the AC conductance as the function of
$\omega/T_K$. Lower panel: Same as a function of $\omega/T^\ast$. 
}
\label{fig:Im_G_NRG}
\end{figure}

Performing a Hilbert transform, 
we  also determined numerically the real part
of the $T$-matrix, $\re t(\omega)$, and from that the imaginary part of the
conductance. These results are shown in Fig.~\ref{fig:Im_G_NRG}. It is 
quite remarkable that, although the scaling is not perfect because of the
insufficient accuracy of the Hilbert transform and the various integrations, 
clearly, the {\em amplitude}  of the  low temperature peak at 
$\omega\sim T^\ast$ does  not change as $T^\ast$ goes to 0. (Note that 
 $T^\ast$ varies over  two orders of magnitudes.) This
behavior is indeed expected based upon Eq.~\eqref{eq:im_cross-over}.
The numerical results confirm that for $J_1>J_2$ and  $J_1<J_2$ the
coefficients $d_\pm$  have different signs,  $d_+> 0$, and
$d_-< 0$, and that $\im G(\omega)$ has a {\em double peak
  structure}: it  has one peak at $\omega\sim T_K$ corresponding to the cross-over to the 
two-channel Kondo fixed point, and also another peak at $\omega\sim T^\ast$
related  to the non-Fermi liquid Fermi liquid cross-over.

It is interesting to observe from Figs.~\ref{fig:T_scaled_doubled},
\ref{fig:G}, and \ref{fig:Im_G_NRG} that the range of two-channel Kondo scaling 
does {\em not} reach from $T_K$ down to the cross-over scale $T^\ast$, but rather it
stops at a much higher energy scale, $\sim \sqrt{T^\ast T_K}$, where  
corrections from the leading relevant operators start to dominate over the 
leading irrelevant operator of the two-channel Kondo fixed point. 

\subsection{Effects of magnetic field}

We also performed calculations for $J_1=J_2$ in the presence 
of a local magnetic field, $B$. As mentioned earlier, a small local 
magnetic field destroys the non-Fermi liquid state and drives the system to 
a trivial, Fermi liquid fixed point below a scale 
$T_B=B^2/T_K \ll T_K$.\cite{AffleckLudwigPangCox}

\begin{figure}[tp]
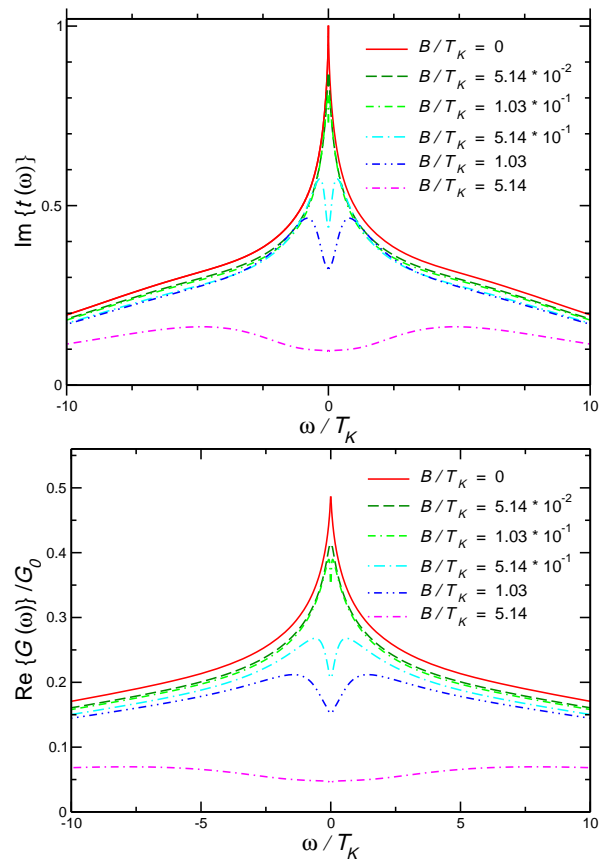

\includegraphics[width=0.9\columnwidth,clip]{T_magn.eps}
\includegraphics[width=0.9\columnwidth,clip]{G_magn.eps}
\caption{Top: Imaginary part of the on-shell $T$-matrix in the presence of a
  magnetic field and no channel asymmetry as the function $\omega/T_K$. 
  Lower curves correspond to higher magnetic fields. Bottom: 
  AC conductance in the presence of a magnetic field and no channel 
  asymmetry as the function of  $\omega/T_K$. 
  Lower curves correspond to higher magnetic field values.
}
\label{fig:T_magn}
\end{figure}

Some typical results are shown in Fig.~\ref{fig:T_magn}. At large magnetic
fields, $B>T_K$, the Kondo resonance is clearly split below the Zeeman field, 
and $\re G(\omega)$ exhibits a dip for $|\omega| < B$. The width 
of this dip gradually decreases as one decreases the size of the field $B$, 
and its depth becomes smaller and smaller. However, it is not clear from the
numerics if there is a critical field value, $B_C$ below which the dip actually 
disappears, as is the case, e.g. for the single-channel Kondo model. 
In fact, the numerical results
 seem to show just the opposite, i.e. that $\re G(\omega)$ remains a
{\em non-monotonous} function in {\em any} finite  magnetic field, 
and only the height
and width of the dip at $\omega\sim T_B$ get smaller and smaller for smaller 
magnetic fields while the dip itself is always present. 
This would indeed naturally follow from a simple scaling argument:
for $B< T_K$ a magnetic energy scale is generated, $T_B = B^2/T_K$, 
and at this energy the real part of the conductance is expected to be 
$\re G(\omega\approx T_B) \approx G_0\;[1/2 -\alpha |B|/T_K]$.   
On the other hand, from Bethe Ansatz\cite{BA} we know the exact phase shifts,
and from that it immediately follows that 
the DC conductance is given by  $G(\omega=0) \approx G_0\;[1/2 - C\;
|B|/T_K\log(T_K/|B|)]$ at $T=0$ temperature, with $C$ a constant of the order
of unity.\cite{LaciGlazman}  This observation suggests that in {\em any}
 finite magnetic field  $G(\omega)$ displays a dip, which has a width 
$\Delta \omega \sim T_B$, and height $\Delta G \sim |B|/T_K\log(T_K/|B|)$.
Similar behavior is expected as a function of temperature, too. 

It is not clear either, if  $G(\omega)$ becomes a universal function 
of $\omega/T_B$. In fact, it has been shown in a special, very anisotropic
limit that no such universal function exists for the non-linear 
DC conductance.\cite{Schiller}  We can argue that the same probably holds 
for the linear AC conductance, although we do not have a rigorous proof.

Unfortunately, from a numerical point of view 
the calculations in a magnetic field turned out to be extremely 
difficult: first of all, for the positive and negative frequency parts of the
spectral function one looses somewhat different amounts  
of spectral weight. This effect turns out to be extremely large in the 
2-channel Kondo case, probably as a consequence of the extreme sensitivity of
the non-Fermi liquid fixed point to the magnetic field. 
Therefore, for a given spin direction,  one needs to {\em match} these positive
and negative-frequency parts  at the origin. Although this is a standard
procedure  followed by most groups, this leads to a large uncertainty in case
of the 2-channel Kondo model. In fact, despite the extensive symmetries 
used, we were not able to obtain data of sufficient accuracy in the most
interesting regime, $\omega\ll T_B=B^2/T_K \ll T_K$, 
even using Hofstetter's  density matrix NRG (DMNRG) method.\cite{Hofstetter} 
Therefore, we were not able to investigate the issue of universal  cross-over 
functions for  $J_1=J_2$ and $T_B=B^2/T_K \ll T_K$. We therefore consider 
these numerical results only as indicative but not decisive.

We also need to recall the well-known fact that NRG produces an artificial
broadening proportional to $\omega$ of the peaks occuring at finite
frequencies. Thus, the correct shape of these split peaks is presumably 
significantly sharper than that shown by the NRG results.

\section{Conclusions}

In this paper,  we have studied the AC transport
properties of a double dot device realized recently by Potok {\em et al.}\
to reach the two-channel Kondo fixed point. First we derived 
an analytical expression for the linear
conductance in the regime where charge fluctuations are small 
and the system can be described by a Kondo Hamiltonian. 
Our formula relates the AC conductance to the eigenvalue $t(\omega)$
of the dimensionless on-shell $T$-matrix, and is  valid at any temperature and
for any frequency. 
Our expression is the  analogue of the formula obtained recently by Sindel
{\em et al.}\ for the Anderson model\cite{sindel05} and  it carries over 
to most Kondo-type Hamiltonians. 

\begin{figure}[bp]
\includegraphics[width=0.9\columnwidth,clip]{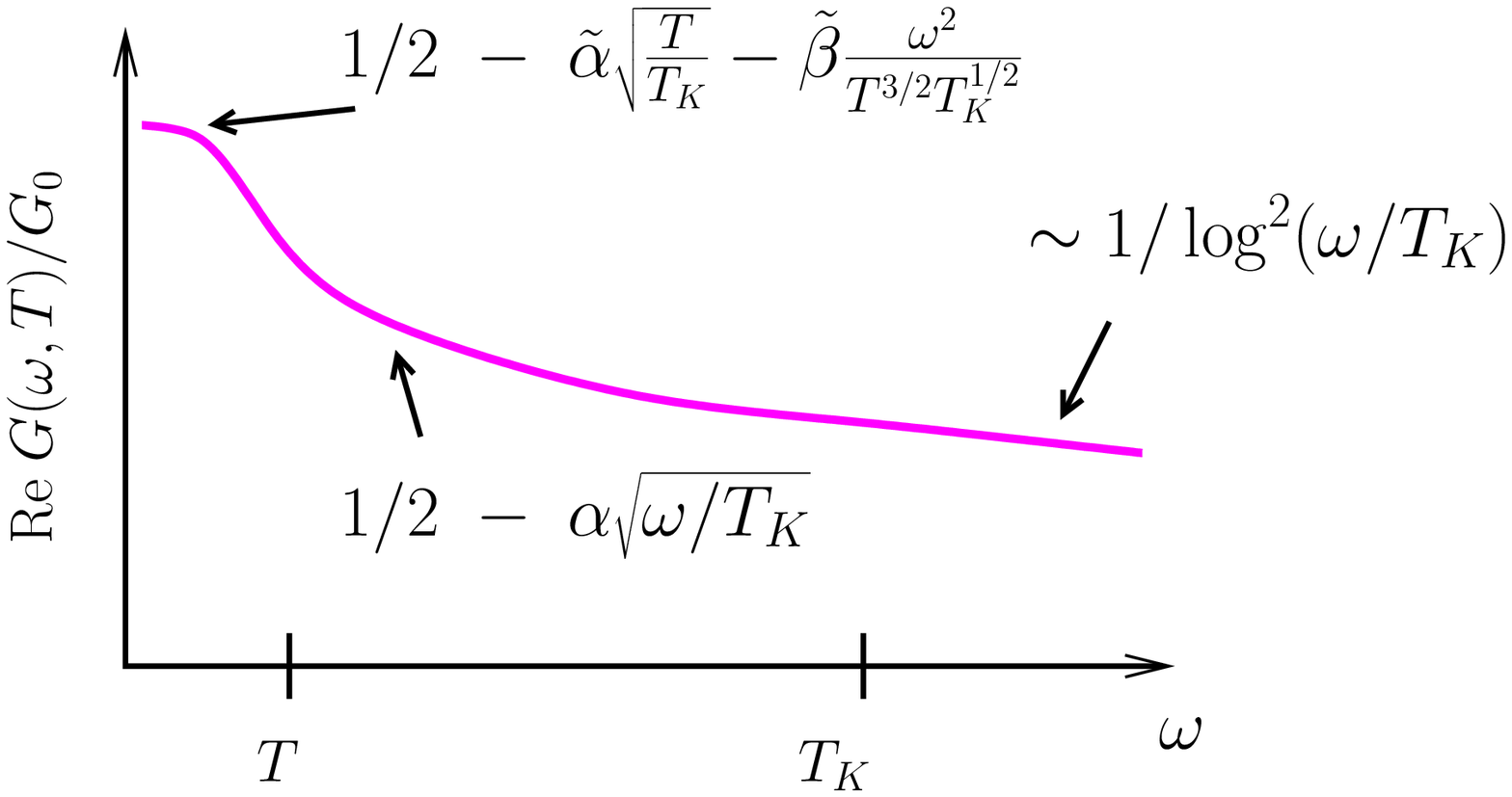}
\includegraphics[width=0.8\columnwidth,clip]{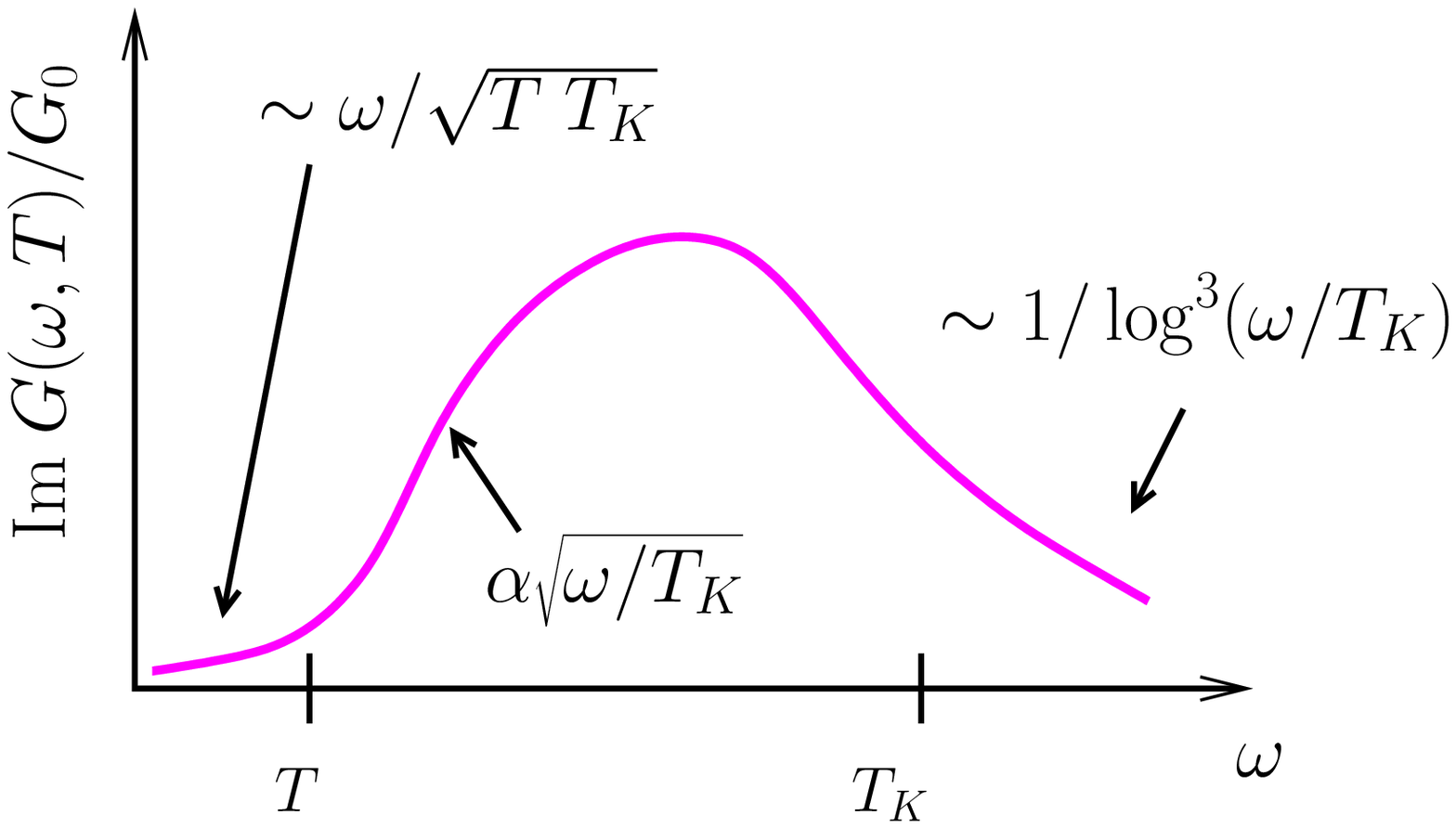}
\caption{Sketch of the real and imaginary parts of the 
 AC conductance for $J_1=J_2$ and $\omega,T\ll T_K$.
}
\label{fig:G_w_T_sketch}  
\end{figure}

The general properties of the two-channel Kondo fixed point, known from 
conformal field theory,  allowed us to make many 
quantitative and qualitative predictions for the AC conductance, $G(\omega)$:
for equal couplings to the two channels, $G(\omega)$ shows an
$\sqrt{\omega/T_K}$ singularity at the two-channel Kondo fixed point.
Using the results of conformal field theory,\cite{AffleckLudwig} we were able to compute the 
real and imaginary parts of the  function $G(\omega,T)$ 
and  determine the corresponding 
scaling functions for both the real and  the
imaginary parts of the conductance through the dot 
in the universal  regime,  $\omega,T\ll  T_K$ and $J_1=J_2$. 
The generic properties of the AC conductance in this regime are summarized
in Fig.~\ref{fig:G_w_T_sketch}.

\begin{figure}[tp]
\includegraphics[width=\columnwidth,clip]{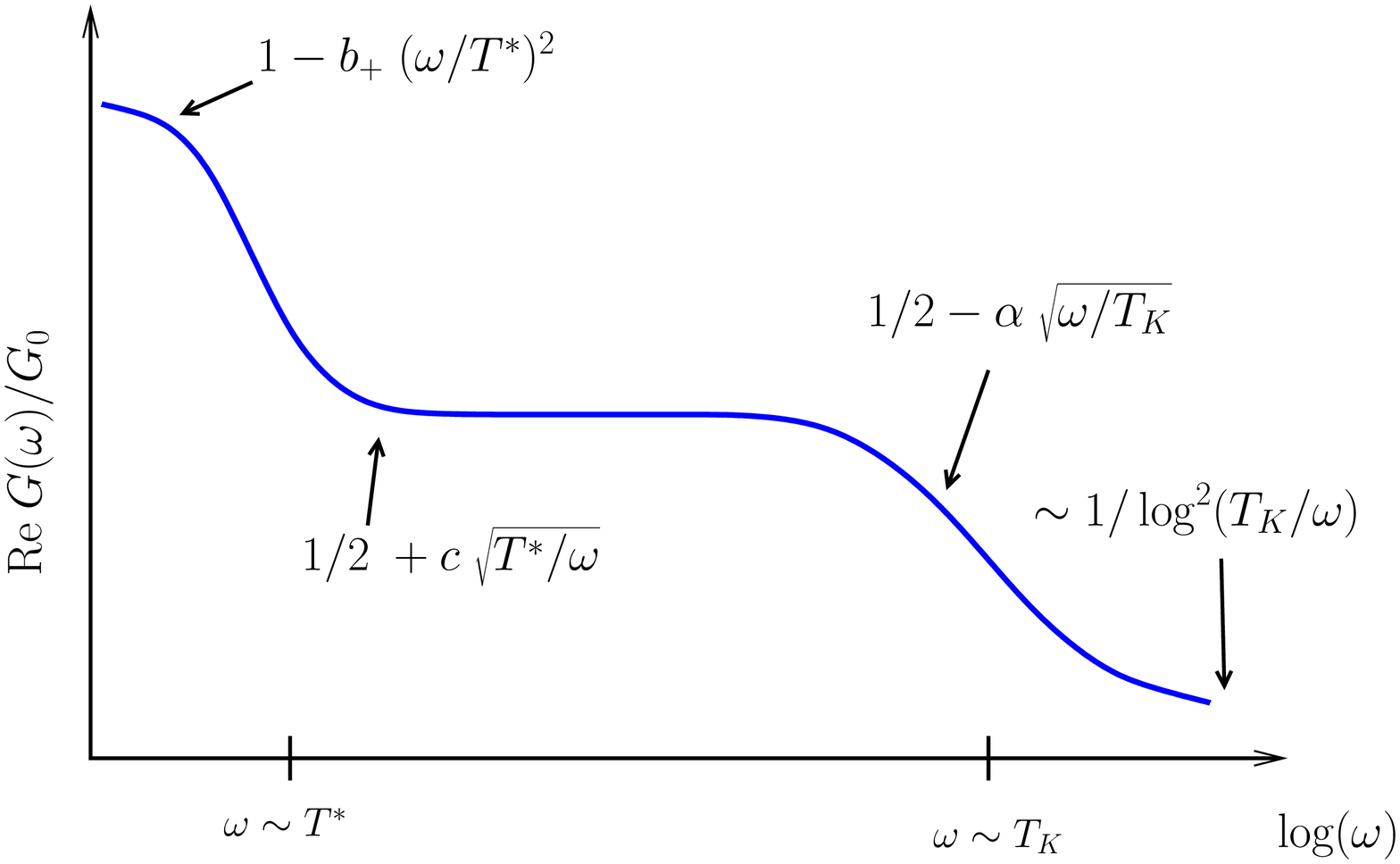}
\includegraphics[width=\columnwidth,clip]{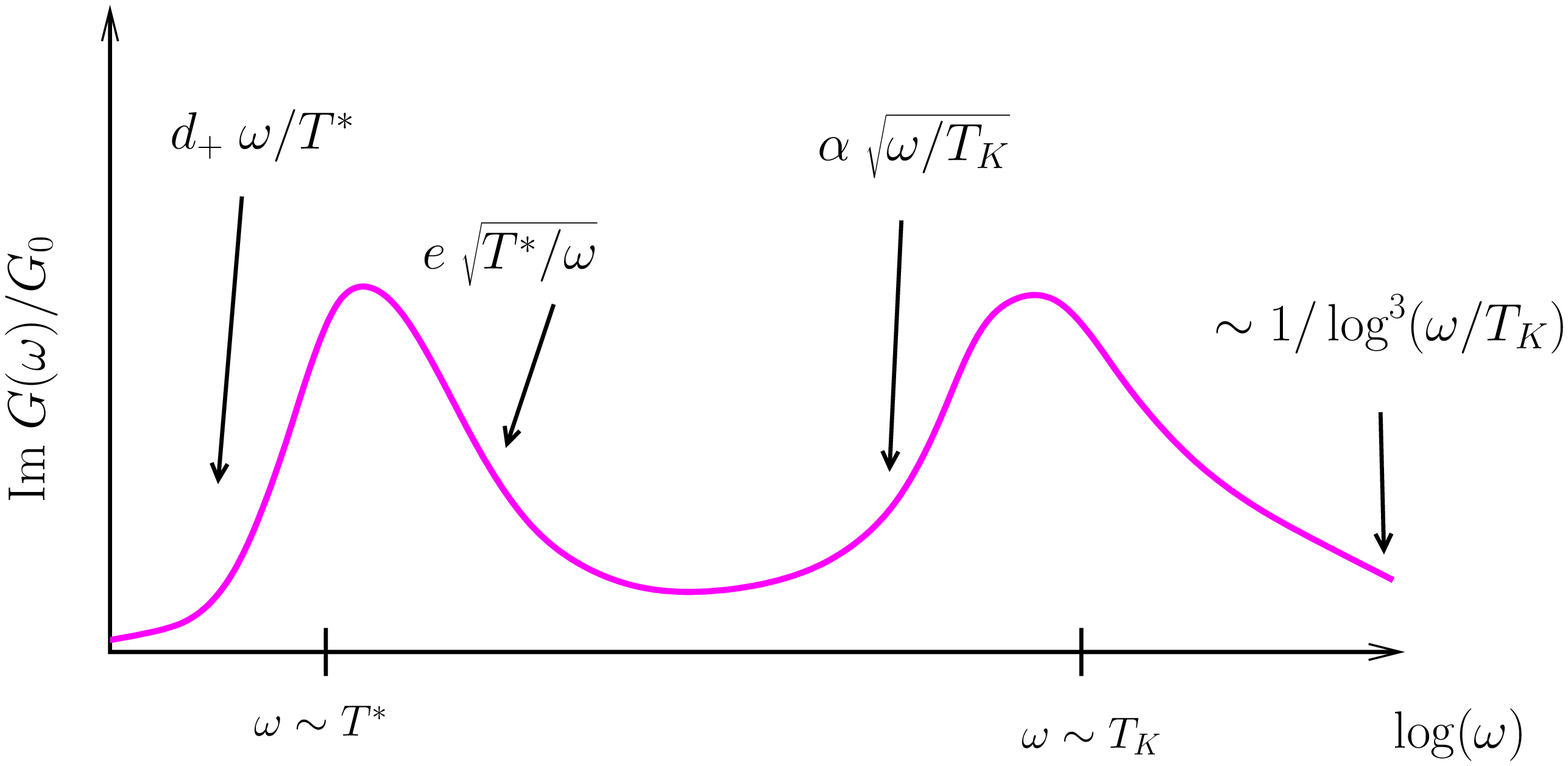}
\caption{Sketch of the real and imaginary parts of the 
 $T=0$ temperature AC conductance for $J_1>J_2$. The various powers 
shown in the picture follow from conformal field theory. The high frequency
behavior is a result of perturbation theory. We assumed electron-hole symmetry.
}
\label{fig:overall_sketch}
\end{figure}

Conformal field theory also gave us a way 
to predict  the basic properties of $\re G(\omega)$ and $\im G(\omega)$
at $T=0$ temperature,  in the presence of channel symmetry-breaking
(see Fig.~\ref{fig:overall_sketch}).
For  $J_1\ne J_2$
$\re G(\omega)$ crosses over to a much smaller or a much larger value (depending on the sign
of asymmetry) at the Fermi-liquid scale $T^\ast$,   
below which it becomes an analytical function of $\omega$.
This cross-over at $\omega \sim T^\ast$ is described by  universal cross-over
functions that we have determined numerically. The asymptotic properties of
the real and imaginary parts of the conductance are dictated 
by conformal field theory (see Eqs.~\eqref{eq:cross-over} and \eqref{eq:im_cross-over}). 
It is quite remarkable
that $\im G(\omega)$ shows a double peak structure at frequencies 
$\omega\sim T^\ast$ and $\omega\sim T_K$. Both peaks are of amplitude
$\sim G_0$ but the  sign of the peak at $\omega\sim T^\ast$ changes with the sign of 
$J_1-J_2$.

One of the important conclusions that one could draw from the 
analysis of $G(\omega)$ was, that the two-channel Kondo regime is, in a sense,
restricted to the regime, $\sqrt{T_K T^\ast} < T, \omega < T_K$: Although it is
true that the entire regime,  $T^\ast < T, \omega < T_K$ is governed by the
2-channel Kondo fixed point, for $ T, \omega < \sqrt{T_K T^\ast} $
the leading relevant operator is more important than the leading irrelevant
operator, and therefore, the scaling curves characteristic to the 
two-channel Kondo fixed point itself cannot be seen in this regime. 
This refines somewhat the phase diagram of the two-channel Kondo model, as already
indicated in Fig.~\ref{fig:conductance}. The two-channel Kondo scaling 
regime is thus limited by a boundary $\sim |J_1 - J_2|$.

We have also investigated the effects of a small Zeeman field 
on the AC conductance. For $B> T_K$ the AC conductance 
exhibits a dip whose width is just $B$. Numerically we find that, apparently, 
this  dip survives for any small magnetic field,  $B< T_K$. 
This would indeed be in agreement with a simple scaling argument 
we presented, that also predicts a similar behavior as a function of 
temperature. In other words, at the two-channel Kondo fixed point the Kondo
resonance appears to be split at {\em any} magnetic field. 
Unfortunately, while our numerics seems to support this picture, it
 is not accurate enough  in  the regime, $B\ll T_K$ to give a decisive answer. 
We remark that the logarithmic magnetic field dependence  of the phase shift  
 would also probably imply that  universal  scaling 
(i.e. $T/T_B$ scaling and the disappearance of the scale 
$T_K$ for $T,T_B\ll T_K$) should be destroyed by logarithmic 
corrections in the presence of magnetic field.

We would like to thank 
%D. Goldhaber-Gordon 
F.\ Anders, A.\ Schiller and L.\ Udvardi for helpful
discussions. This research has been supported by Hungarian grants
OTKA Nos. NF061726, T046267, T046303, D048665, by the DFG center for 
functional nanostructures (CFN),  and by Sonderforschungsbereich 631. 
G.\ Z.\ acknowledges the hospitality of the CAS, Oslo, and L.\ B.\ the financial
support received from the Bolyai Foundation. 

%\begin{figure}[tb]
%\includegraphics[width=7cm]{qcp_sketch.eps}
%\caption{Sketch of the phase diagram of the two-channel Kondo model.}
%\label{fig:phasediag}
%\end{figure}

\end{document}